%% file: main.tex
\documentclass[sigconf]{acmart}
\usepackage{makecell}
\usepackage{multirow}

\AtBeginDocument{%
  \providecommand\BibTeX{{%
    \normalfont B\kern-0.5em{\scshape i\kern-0.25em b}\kern-0.8em\TeX}}}

\copyrightyear{2025}
\acmYear{2025}
\setcopyright{acmlicensed}\acmConference[KDD '25]{Proceedings of the 31st ACM SIGKDD International Conference on Knowledge Discovery \& Data Mining}{August 3--7, 2025}{Toronto, Canada}
\acmBooktitle{Proceedings of the 31st ACM SIGKDD International Conference on Knowledge Discovery \& Data Mining (KDD '25), August 3--7, 2025, Toronto, Canada}
\acmPrice{15.00}
\acmDOI{XXXXXXX.XXXXXXX}
\acmISBN{XXX}

\acmSubmissionID{322}

\usepackage{amsthm}
\newtheorem{definition}{Definition}
\newtheorem{theorem}{Theorem}
\newtheorem{lemma}{Lemma}
\newtheorem{corollary}{Corollary}

\newcommand{\ie}{\emph{i.e.}}
\renewcommand{\shortauthors}{Ruitao Zhu, et al.}

\newcommand{\et}{\emph{et al.}}

\begin{document}

\title{Contextual Generative Auction with Permutation-level Externalities for Online Advertising}

\author{Ruitao Zhu}
\affiliation{
	\institution{Shanghai Jiao Tong University}
    \city{Shanghai}
	\country{China}
}
\email{sjtu_zrt@sjtu.edu.cn}

\author{Yangsu Liu \\ Dagui Chen \\ Zhenjia Ma}
\affiliation{
	\institution{Alibaba Group}
    \city{Beijing}
	\country{China}
}
\email{liuyangsu.lys@taobao.com}
\email{dagui.cdg@taobao.com}
\email{mazhenjia.mzj@alibaba-inc.com}

% \author{Dagui Chen}
% \affiliation{
% 	\institution{Alibaba Group}
%     \city{Beijing}
% 	\country{China}
% }
% \email{dagui.cdg@taobao.com}

% \author{Zhenjia Ma}
% \affiliation{
% 	\institution{Alibaba Group}
%     \city{Beijing}
% 	\country{China}
% }
% \email{mazhenjia.mzj@alibaba-inc.com}

\author{Chufeng Shi}
\affiliation{
	\institution{Shanghai Jiao Tong University}
    \city{Shanghai}
	\country{China}
}
\email{cfshi99@sjtu.edu.cn}

\author{Zhenzhe Zheng}
\authornote{Corresponding author.}
\affiliation{
	\institution{Shanghai Jiao Tong University}
	\city{Shanghai}
	\country{China}
}
\email{zhengzhenzhe@sjtu.edu.cn}

\author{Jie Zhang \\ Jian Xu \\ Bo Zheng}
\affiliation{
	\institution{Alibaba Group}
    \city{Beijing}
	\country{China}
}
\email{kongpan.zj@taobao.com}
\email{xiyu.xj@alibaba-inc.com}
\email{bozheng@alibaba-inc.com}

% \author{Jian Xu}
% \affiliation{
% 	\institution{Alibaba Group}
%     \city{Beijing}
% 	\country{China}
% }
% \email{xiyu.xj@alibaba-inc.com}

% \author{Bo Zheng}
% \affiliation{
% 	\institution{Alibaba Group}
%     \city{Beijing}
% 	\country{China}
% }
% \email{bozheng@alibaba-inc.com}

\author{Fan Wu}
\affiliation{
	\institution{Shanghai Jiao Tong University}
	\city{Shanghai}
	\country{China}
}
\email{fwu@cs.sjtu.edu.cn}

\renewcommand{\shortauthors}{Ruitao Zhu, et al.}

\begin{abstract}
\input{sections/Abstract}
\end{abstract}




\maketitle
\allowdisplaybreaks

\input{sections/Introduction}

\input{sections/Preliminaries}

\input{sections/Methodology}

\input{sections/Experiments}

\input{sections/Related_work}

\input{sections/Conclusion}

\begin{acks}
This work was supported in part by National Key R\&D Program of China (No. 2023YFB4502400), in part by China NSF grant No. 62322206, 62132018, 62025204, U2268204, 62272307, 62372296, in part by Alibaba Group through Alibaba Innovative Research Program. The opinions, findings, conclusions, and recommendations expressed in this paper are those of the authors and do not necessarily reflect the views of the funding agencies or the government.
\end{acks}

\bibliographystyle{ACM-Reference-Format}
\bibliography{ref}

% \newpage
\input{sections/Appendix}

\end{document}

%% file: sections/Abstract.tex
Online advertising has become a core revenue driver for the internet industry, with ad auctions playing a crucial role in ensuring platform revenue and advertiser incentives. Traditional auction mechanisms, like GSP, rely on the independent CTR assumption and fail to account for the influence of other displayed items, termed externalities. Recent advancements in learning-based auctions have enhanced the encoding of high-dimensional contextual features. However, existing methods are constrained by the ``allocation-after-prediction'' design paradigm, which models set-level externalities within candidate ads and fails to consider the sequential context of the final allocation, leading to suboptimal results. This paper introduces the Contextual Generative Auction (CGA), a novel framework that incorporates permutation-level externalities in multi-slot ad auctions. Built on the structure of our theoretically derived optimal solution, CGA decouples the optimization of allocation and payment. We construct an autoregressive generative model for allocation and reformulate the incentive compatibility (IC) constraint into minimizing ex-post regret that supports gradient computation, enabling end-to-end learning of the optimal payment rule. Extensive offline and online experiments demonstrate that CGA significantly enhances platform revenue and CTR compared to existing methods, while effectively approximating the optimal auction with nearly maximal revenue and minimal regret.

%% file: sections/Introduction.tex
\section{Introduction}

Online advertising serves as a cost-efficient and precise channel for advertisers to promote contents to millions of online users, which has become the main revenue source for the internet industry \cite{edelman2007internet, jansen2008sponsored}. 
Upon receiving a user request, online platforms conduct real-time auctions to determine ad allocation across multiple slots on a webpage and compute the payment of each advertiser obtaining an ad slot.
The optimal ad auction aims to maximize platform revenue while satisfying the economic properties, such as incentive compatibility (IC) and individual rationality (IR), which will be detailed in Section \ref{section_preliminaries}, along with computational complexity constraints for online deployment.

Traditional ad auctions jointly consider advertisers' bids and ads' click-through rate (CTR). The Generalized Second Price (GSP) auction has long been the benchmark for ad auctions due to its interpretability and ease of implementation, and has further evolved into variants such as ugsp \cite{ugsp} and DeepGSP \cite{DeepGSP}.
% , and DNA \cite{dna} through the integration of deep learning techniques.
Although existing efforts have demonstrated remarkable performance, the separated CTR assumption employed by GSP, which posits that the users' clicking depends solely on the ad itself, restricts the model's predictive power by overlooking other displayed items.
In reality, the results page presented to users contains multiple items, and users compare factors such as prices and appearances before making decisions; therefore, other displayed items will influence the target ad's CTR \cite{extr, zhang2021constructing}.
From the perspective of mechanism design, the effect of external items is defined as \textit{externalities} \cite{externalities_1} in online advertising. The empirical study on user behaviors \cite{externalities_2} also indicates that optimal ad auction design must take externalities into account.

Recently, learning-based auctions such as Deep Neural Auction (DNA) \cite{dna} and Score Weighted VCG (SW-VCG) \cite{li2023learning} are proposed to better capture externalities and enhance platform revenue. However, these methods are limited by the "allocation-after-prediction" design paradigm, as the prediction process remains agnostic to the context within the allocation results.
In essence, these methods only model information within the candidate ad set (\textit{set-level externalities}), failing to incorporate the sequential context that affects the CTR of each ad within the final allocation (\textit{permutation-level externalities}), resulting in suboptimal allocation.
Studies on reranking \cite{pei2019personalized, xi2021context, ren2024non} in recommendation systems similarly consider item correlations within the displayed list. Nonetheless, their focus on enhancing overall user feedback neglects advertiser strategic behaviors and thus fails to incorporate economic constraints like IC and IR, hindering their applicability to revenue maximization in online auction environments.

\begin{figure}[tp]
    \centering
    \includegraphics[width=\linewidth]{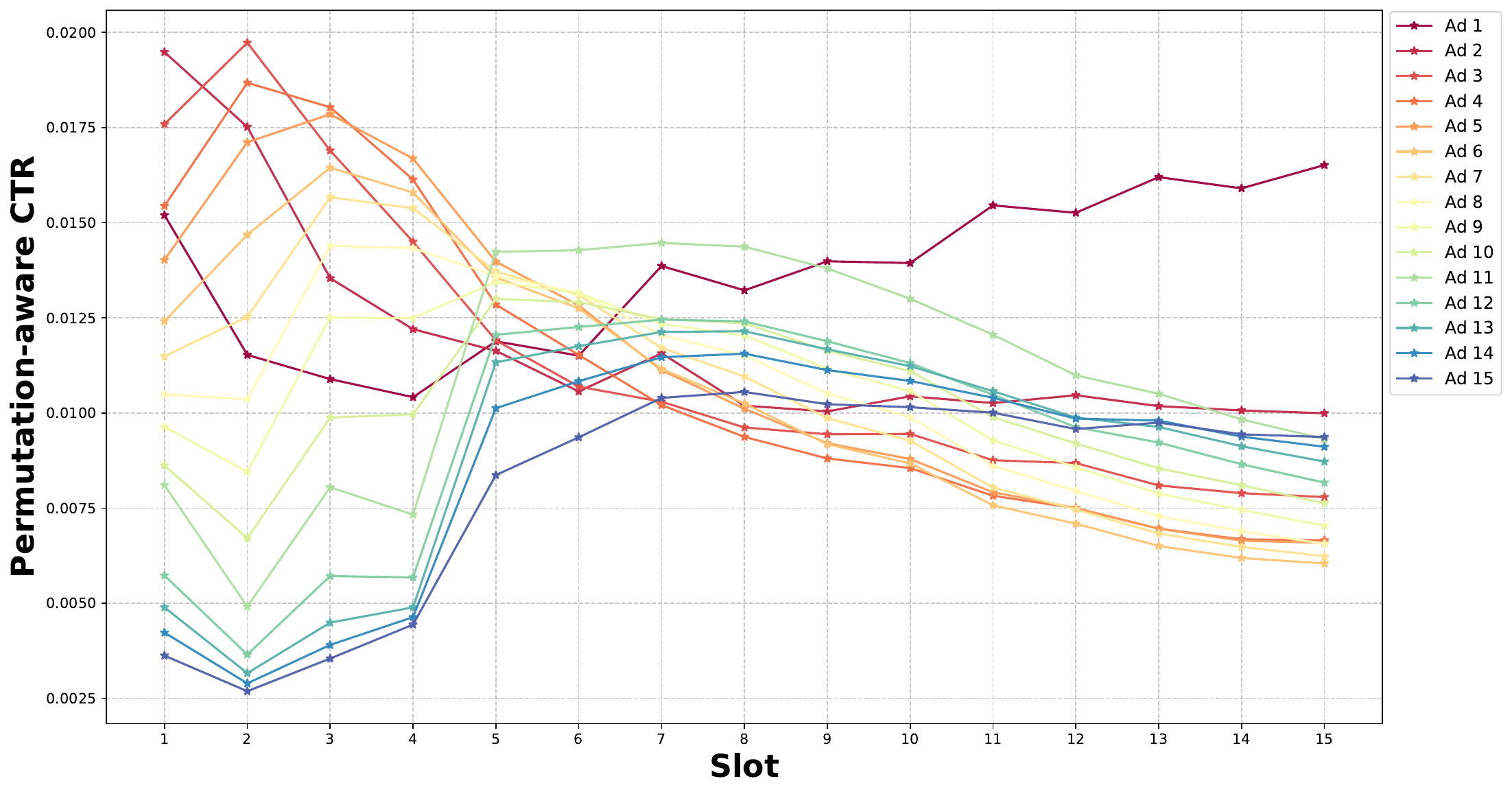}
    \caption{The Permutation-aware ad CTR as a function of slot on \textit{Taobao} is non-monotonic.
       }
    \label{figure: monotone_ctr}
\end{figure}

Designing optimal auction with \textit{permutation-level externalities} for online ad platform faces three critical challenges:
(\textbf{\romannumeral1}) The existing ``allocation-after-prediction'' design paradigm cannot perceive the sequential context while predicting ad value.
(\textbf{\romannumeral2}) Employing a VCG-like approach to traverse all permutations can achieve optimal allocation, but the high computational complexity makes it infeasible for online applications.
The challenge lies in the efficient exploration of optimal sequences within the factorial-level permutation space.
(\textbf{\romannumeral3})
The IC condition requires that each advertiser's expected value is non-decreasing with her bid \cite{myerson1981optimal}. While most existing methods ensure that a higher bid secures the same or a higher slot for the advertiser, permutation-level externalities can cause higher slots to yield lower CTRs. As illustrated in Figure \ref{figure: monotone_ctr}, industrial data from the \textit{Taobao} platform shows that the permutation-level CTR of ads is not monotonic with respect to their allocated slot. Consequently, designing an ad auction with an IC constraint under permutation-level externalities is a non-trivial problem.

To address the above limitations, this paper proposes the Contextual Generative Auction (CGA), designed to optimize platform revenue with guaranteed economic properties. 
The framework of CGA adheres to the structure of our theoretically derived optimal DSIC (dominant strategy incentive compatible) auction, which decouples the optimization of allocation and payment. CGA employs the Generator-Evaluator (G-E) paradigm \cite{GRN} to capture permutation-level externalities. Specifically, the Generator utilizes the autoregressive approach to generate the ad sequence as allocation, guided by the well-trained Evaluator that refines contextual interaction within the ad sequence. Additionally, directly calculating the optimal payment rule is infeasible, since the allocation rule of CGA is implemented via a generative model. Inspired by works on multi-item auction design that uses neural networks to effectively recover analytical solutions \cite{dutting2019optimal}, we restate the IC constraint to require zero expected ex-post regret for ad auction and learn the optimal payment by pushing gradients through the regret term.
Our main contributions are summarized as follows:

$\bullet$ We formulate the multi-slot auction with permutation-level externalities and theoretically derive the optimal DSIC auction\footnote{Li ~\et~ \cite{li2023learning} directly rewrites Myerson Auction into an externality-aware form, yielding similar conclusions. However, when incorporating permutation-level externalities, unlike traditional single-item auctions, the ad CTR is influenced by the allocation outcome. This impact complicates the transition of the monotone allocation condition, which satisfies IC constraint, from Myerson Auction. Consequently, the preservation of optimality is non-trivial. We discuss this in detail in Lemma \ref{lemma: monotone allocation} and Appendix \ref{appendix: proof of monotone allocation}.}, providing theoretical guarantees for generative auction design.

$\bullet$ To model the permutation-level externalities, we break the "allocation-after-prediction" design paradigm of learning-based auctions by introducing an autoregressive generative model that directly generates allocations. Building on the structure of the optimal mechanism, the proposed CGA leverages the G-E paradigm to optimize allocation and minimizes differentiable ex-post regret to learn the optimal payment rule.

$\bullet$ Experiments on large-scale industrial data and online A/B tests demonstrate that CGA outperforms existing methods in platform revenue and CTR, and effectively approximates the optimal DSIC auction with nearly maximal revenue and negligible regret.

%% file: sections/Preliminaries.tex
\section{PRELIMINARIES}
\label{section_preliminaries}

\subsection{Multi-slot Auctions with Permutation-level Externalities}
\textbf{Multi-slot auction.}
The ad auction for online advertising can be abstracted as a typical multi-slot auction design problem. 
Formally, when a user request arrives, there are $n$ advertisers\footnote{Advertisers can either manually adjust bids or employ the platform's auto-bidding agent \cite{deng2021towards, balseiro2021landscape, xing2023truthful} for automated bidding.}
bidding for $k$ ($k\leq n$) ad slots with each advertiser $i$ submits a bid $b_i$ based on her private value $v_i$ for a click to $ad_i$\footnote{We consider the single-parameter and Cost-per-Click (CPC) auction setting, \ie, each advertiser submits a bid and gets paid for a click event, which aligns with the practical scenarios of online ad platforms and remains consistent with related work \cite{dna, li2023learning, nma}.}.
Advertiser $i$'s private value is drawn independently from a distribution $f_i(\cdot)$, and $F_i(\cdot)$ denotes the cumulative distribution function (cdf) of the probability density function (pdf) $f_i(\cdot)$. Let $\boldsymbol{v}=(v_1, v_2, \cdots, v_n)$ denote the value profile for all advertisers, with $\boldsymbol{v_{-i}}$ representing the value profile excluding the element $v_i$, similarly for $\boldsymbol{b}$ and $\boldsymbol{b_{-i}}$. The auctioneer, \ie, the ad platform, knows the distributions $\boldsymbol{f}=(f_1, f_2, \cdots, f_n)$ (derived from historical data), but does not know the true value profile $\boldsymbol{v}$.

Given bid profile $\boldsymbol{b}$, user feature vector $\boldsymbol{u}$, and the collection of all ad feature vectors $\boldsymbol{X}=(\boldsymbol{x_1}, \boldsymbol{x_2}, \cdots, \boldsymbol{x_n})$, the ad auction mechanism is then formalized as $\mathcal{M}\langle \mathcal{A}(\boldsymbol{b}; \boldsymbol{X}, \boldsymbol{u}),\mathcal{P}(\boldsymbol{b}; \boldsymbol{X}, \boldsymbol{u})\rangle$. The allocation rule $\mathcal{A}$ decides the ad-slot allocation, represented by the allocation matrix $\boldsymbol{A}=(a_{ij})_{n\times k} \in \{0,1\}^{n\times k}$, where $a_{ij}=1$ means that ad $i$ is allocated to slot $j$ and $a_{ij}=0$ otherwise\footnote{For notational simplicity, we slightly abuse matrix notation and rewrite $\boldsymbol{A}=(a_{A_1}, a_{A_2}, \cdots, a_{A_k})$ to denote the sequential allocation of ads to $k$ slots, where $A_i \in [n]$ denotes the index of ad allocated to the $i$-th slot of allocation $\boldsymbol{A}$.}. Note that an ad is allocated to at most one slot, and each slot must be assigned one ad, hence a feasible allocation matrix $\boldsymbol{A}$ satisfies $\sum_{j=1}^k a_{ij}\leq 1, \forall i$ and $\sum_{i=1}^n a_{ij}=1, \forall j$;
The payment rule $\mathcal{P}$ decides the payment of each ad, represented by the payment vector $\boldsymbol{p}=(p_{i})_n \in \mathbb{R}^{n}$.

\textbf{Permutation-level externalities.}
The externalities in ad auctions mean that the winning ads' utility is also influenced by other winning ads \cite{externalities_1}.
For multi-slot ad auctions, the externality effect is reflected in the CTR of ads.
Specifically, CTR models can be abstracted as functions mapping from ad features and user features to the probability of a user clicking an ad. 
We further consider a permutation-aware CTR model $\Theta$, which takes allocation $\boldsymbol{A}$ and associated features as input, and captures the inter-ad influence within $\boldsymbol{A}$.
Note that model $\Theta$ is permutation-aware, meaning that even when two allocations contain the same set of ads, such as $\boldsymbol{A_1}=(ad_1, ad_2, ad_3)$ and $\boldsymbol{A_2}=(ad_3, ad_2, ad_1)$ with $k=3$, the differing permutations result in a variance in the external impact on the CTR of $ad_2$.
Formally, we denote the CTR of $ad_i$ as $\Theta(\boldsymbol{x_i}; \boldsymbol{A}, \boldsymbol{X}, \boldsymbol{u})$.
For ease of notation, let $\Theta_i(\boldsymbol{b};\boldsymbol{X}, \boldsymbol{u})=\Theta(\boldsymbol{x_i}; \mathcal{A}(\boldsymbol{b}; \boldsymbol{X}, \boldsymbol{u}), \boldsymbol{X}, \boldsymbol{u})$.
The permutation-level externalities is encoded in the mapping process from allocation $\boldsymbol{A}$ to CTR via model $\Theta$.

\textbf{Problem formulation.}
Given the mechanism $\mathcal{M}\langle \mathcal{A},\mathcal{P}\rangle$, the expected utility of an advertiser with valuation $v_i$ is given by:
\begin{equation*}
  \label{equ: utility}
  u_i(v_i;\boldsymbol{b};\boldsymbol{X}, \boldsymbol{u}) = (v_i - p_i(\boldsymbol{b};\boldsymbol{X}, \boldsymbol{u}) )\cdot \Theta_i( \boldsymbol{b};\boldsymbol{X}, \boldsymbol{u}).
\end{equation*}
Considering that advertisers can employ strategies aimed at maximizing their utilities through misreporting their values, we introduce two essential properties of ad auction for the stability of the advertising platform: \textit{dominant strategy incentive compatible} (DSIC, or IC) and \textit{individually rational} (IR).

\begin{definition}
  \label{IC}
  An auction is DSIC or IC, if each advertiser's utility is maximized by reporting truthfully no matter what the other advertisers report. Formally, 
  \begin{equation*}
    u_i(v_i;v_i, \boldsymbol{b_{-i}};\boldsymbol{X}, \boldsymbol{u}) \geq u_i(v_i;b_i, \boldsymbol{b_{-i}};\boldsymbol{X}, \boldsymbol{u}), \forall i \in [n], \forall v_i,b_i \in \mathbb{R^+}.
  \end{equation*}
\end{definition}

\begin{definition}
  \label{IR}
  An auction is IR, if every advertiser’s payment does not exceed her submitted bid. Formally,
  \begin{equation*}
    p_i(\boldsymbol{b}; \boldsymbol{X}, \boldsymbol{u}) \leq b_i, \forall i \in [n].
  \end{equation*}
\end{definition}

The goal is to find an auction $\mathcal{M}$ that maximizes the expected revenue of ad platform:
\begin{equation*}
  \mathbb{E}_{\boldsymbol{v} \sim \boldsymbol{F}} Rev^{\mathcal{M}}(\boldsymbol{b},\boldsymbol{X}, \boldsymbol{u}) = \sum_{i=1}^n p_i(\boldsymbol{b}; \boldsymbol{X}, \boldsymbol{u}) \Theta_i(\boldsymbol{b};\boldsymbol{X}, \boldsymbol{u}),
\end{equation*}
and satisfies IC and IR constraints, which can be formulated as:
\begin{equation}
  \label{equ: problem formulation}
  \max_{\mathcal{M}\langle \mathcal{A},\mathcal{P}\rangle} \mathbb{E}_{\boldsymbol{v} \sim \boldsymbol{F}}\  Rev^{\mathcal{M}}(\boldsymbol{b},\boldsymbol{X}, \boldsymbol{u}), \ s.t. \ \text{IC and IR constraints}.
\end{equation}

\subsection{Optimal Multi-slot Auction}
\label{section: theoretical optimality}
To solve the above revenue maximizing auction design problem with IC and IR constraints, an intuitive method is to follow the well-known Myerson auction \cite{myerson1981optimal}.
Myerson auction can be adapted to the setting with permutation-level externalities as follows:

\begin{definition}
  \label{def: optimal mechanism}
  (Myerson auction with permutation-level externalities):
  \begin{itemize}
    \item \textbf{Allocation}: $\mathcal{A} \in argmax_{A}\sum_{i=A_1}^{A_k} \tilde{\phi}(b_i, F_i)  \Theta(\boldsymbol{x_i}; \boldsymbol{A}, \boldsymbol{X}, \boldsymbol{u})$;
    \item \textbf{Payment}: 
    $p_i=\begin{cases} b_i - \frac{\int_0^{b_i}\Theta_i(t, \boldsymbol{b_{-i}};\boldsymbol{X}, \boldsymbol{u})dt}{\Theta_i( \boldsymbol{b};\boldsymbol{X}, \boldsymbol{u})} & \text { if } \Theta_i( \boldsymbol{b};\boldsymbol{X}, \boldsymbol{u})>0 ;\\ 0, & \text { otherwise, }\end{cases} $
  \end{itemize}
  where $\tilde{\phi}(v, F_i)$ denotes the ironed virtual value function \cite{myerson1981optimal}, which is monotone in $v$. 
\end{definition}

Recall Myerson's Lemma \cite{myerson1981optimal}, as formulated in Theorem \ref{lemma: myerson lemma}.
Note that introducing permutation-level externalities diverges from traditional Myerson auction by influencing each ad's CTR based on allocation outcomes. As such, an increase in $ad_i$'s bid does not necessarily increase its CTR due to inter-ad influences within the allocation results. 
This could potentially violate the monotonic allocation requirement of Theorem \ref{lemma: myerson lemma}.
We address this by proving that, even with permutation-level externalities, the monotonic allocation condition in Myerson's Lemma holds if the allocation rule maximizes virtual welfare, as demonstrated in Lemma \ref{lemma: monotone allocation}.
We put the proof in Appendix \ref{appendix: proof of monotone allocation} due to space limitations.

\begin{theorem}
  \label{lemma: myerson lemma}
  (Myerson's Lemma \cite{myerson1981optimal}). For a single-parameter environment, an allocation rule $\mathcal{A}$ is \textit{implementable} if there exists a payment rule $\mathcal{P}$ such the mechanism $\mathcal{M}\langle \mathcal{A},\mathcal{P}\rangle$ is DSIC. The following two claims hold: (1) An allocation rule is implementable if and only if it is monotone. (2) If allocation rule $\mathcal{A}$ is monotone, then the exists a unique payment rule $\mathcal{P}$ such that the mechanism $\mathcal{M}\langle \mathcal{A},\mathcal{P}\rangle$ is DSIC. It is given by:
  \begin{equation*}
    \mathcal{P}_i(b_i, \boldsymbol{b_{-i}}) = b_i \mathcal{A}_i(b_i, \boldsymbol{b_{-i}}) - \int_0^{b_i}\mathcal{A}_i(t, \boldsymbol{b_{-i}})dt.
  \end{equation*}
\end{theorem}

\begin{lemma}
  \label{lemma: monotone allocation}
  (Monotonic allocation with permutation-level externalities). For every $ad_i$ and $\boldsymbol{b_{-i}}$, the obtained CTR $\Theta(\boldsymbol{x_i}; \mathcal{A}(b_i, \boldsymbol{b_{-i}}), \boldsymbol{X}, \\ \boldsymbol{u})$ is nondecreasing in its bid $b_i$ (or we say $\mathcal{A}$ is monotone), if the allocation rule $\mathcal{A}$ maximizes the expected virtual welfare.
\end{lemma}

\begin{theorem}
  \label{theorem: expected virtual welfare}
  For a single-parameter environment, maximizing expected revenue over the space of DSIC auctions is equal to maximizing expected virtual welfare \cite{myerson1981optimal}.
\end{theorem}

Based on Theorem \ref{lemma: myerson lemma} and \ref{theorem: expected virtual welfare} and Lemma \ref{lemma: monotone allocation}, we can deduce the corollary that Myerson auction with permutation-level externalities in Definition \ref{def: optimal mechanism} constitutes the optimal solution to Problem (\ref{equ: problem formulation}).

\begin{corollary}
  \label{corollary: optimal solution}
  The ad auction $\mathcal{M}$, characterized by allocation rule $\mathcal{A}$ and payment rule $\mathcal{P}$ in Definition \ref{def: optimal mechanism}, represents the optimal mechanism with permutation-level externalities that maximizes the platform's expected revenue while satisfying IC and IR constraints. 
\end{corollary}

\subsection{Auction Design as a Learning-based Problem}
A direct approach to implement the allocation rule $\mathcal{A}$, as outlined in Definition \ref{def: optimal mechanism}, involves enumerating all permutations to select the one maximizing virtual welfare, with a computational complexity of $P(n,k)=\frac{n!}{(n-k)!}$. However, for online advertising, taking \textit{Taobao} as an example, where $n \approx 50$ and $k \approx 5$, the high computational complexity fails to meet online real-time inference requirements. Therefore, we parameterize the auction mechanism as $\mathcal{M}^w=\langle\mathcal{A}^{w_a}, \mathcal{P}^{w_p}\rangle$ with parameters $w_a$ and $w_p$, 
and solves a learning problem to determine these parameters.

To impose IC constraint on learning-based auctions and ensure the differentiability of the optimization process, similar to the original work of learning-based multi-item auction \cite{dutting2019optimal}, we introduce the concept of \textit{ex-post regret} for each advertiser to quantify the extent of deviation from IC conditions.
Specifically, the \textit{ex-post regret} for $ad_i$ is defined as the maximum increase in utility that can be obtained through misreporting $b_i'$:
\begin{equation*}
  \label{equ: ex-post regret}
  rgt_i(v_i, \boldsymbol{X}, \boldsymbol{u}) = \max_{b_i'} \{u_i(v_i;b_i',\boldsymbol{b_{-i}};\boldsymbol{X}, \boldsymbol{u})- u_i(v_i;\boldsymbol{b};\boldsymbol{X}, \boldsymbol{u}) \}.
\end{equation*}
Hence, the IC constraint is satisfied if and only if the ex-post regret for each $ad_i$ is zero.
Given $L$ valuation samples from distribution $\mathbf{F}$, we obtain the empirical ex-post regret for $ad_i$: 
\begin{equation}
  \label{equ: empirical ex-post regret}
  \widehat{rgt}_i = \frac{1}{L}\sum_{l=1}^{L} rgt_i(v_i^l, \boldsymbol{X}, \boldsymbol{u}).
\end{equation}  
Then we reformulate the auction design problem (\ref{equ: problem formulation}) as minimizing the expected negated revenue, subject to the constraint that the empirical ex-post regret for each $ad_i$ is zero:
\begin{equation}
  \label{equ: learning-based problem formulation}
  \min_{w} -\mathbb{E}_{\boldsymbol{v} \sim \boldsymbol{F}} Rev^{\mathcal{M}^w}, \ s.t. \ \  \widehat{rgt}_i=0, \ \forall i\in [n].
\end{equation}

%% file: sections/Methodology.tex
\section{Methodology}
\label{section: methodology}
\begin{figure*}[htbp]
  \centering
  \includegraphics[width=0.9\linewidth]{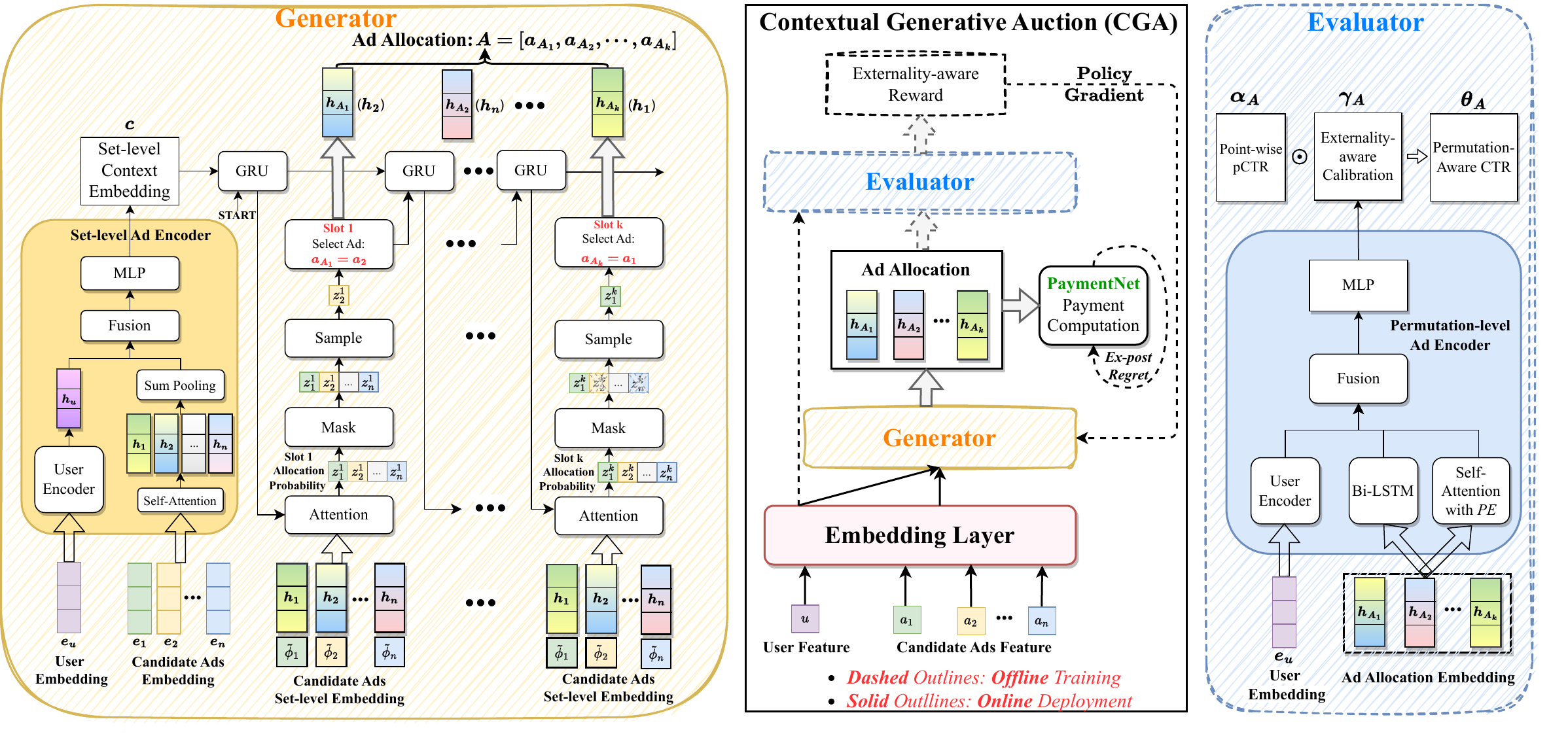}
  \caption{The architecture of Contextual Generative Auction (CGA). The middle part shows the overall framework of CGA, with dashed outlines and arrows depicting offline training components, and solid lines representing online inference paths. The other two parts provide specific implementations of Generator and Evaluator.
     }
    \label{figure: CGA}
\end{figure*}

This section provides a detailed discussion of the Contextual Generative Auction (CGA).
To overcome the limitation of the ``allocation-after-prediction'' paradigm in capturing permutation-level externalities, CGA employs a G-E architecture, as depicted in Figure \ref{figure: CGA}. The Generator leverages an autoregressive module, perceiving the established preceding context to generate an ad sequence. The Evaluator estimates permutation-aware reward within the whole ad sequence, guiding Generator towards optimal allocation through policy gradient. For online inference, only Generator is deployed, ensuring minimal computational latency during ad allocation.
Moreover, a dedicated PaymentNet module is introduced to learn the optimal payment rule, trained via optimizing differentiable ex-post regret.

\subsection{Generator: Autoregressive Generative Module}
\label{section: generator}
According to Theorem \ref{theorem: expected virtual welfare} and Corollary \ref{corollary: optimal solution},
the objective of Generator is to generate an ad sequence $\boldsymbol{A}$ of length $k$ from $n$ candidate ads that maximizes the expected virtual welfare.
We develop an autoregressive generative module, comprising two key components: a permutation-invariant set-level encoder and a permutation-equivariant decoder.
\textit{Permutation invariance} \cite{deepsets}, an architectural property wherein the output remains agnostic to permutations of the inputs, has been shown in DNA \cite{dna} to enhance the platform revenue of ad auctions. \textit{Permutation equivariance} \cite{deepsets} means that any permutation of inputs will result in an identical permutation in the outputs. This property is widely adopted in automated auction design \cite{rahme2021permutation, duan2024scalable, ivanov2022optimal}, with Qin ~\et~ \cite{qin2022benefits} demonstrating its efficacy in improving the generalization ability of learning-based auctions.

\textbf{Permutaion-invariant set-level encoder.}
The ad encoder aims to enhance each ad's representation by modeling set-level externalities.
It encodes the candidate ad set and provides the context embedding $\boldsymbol{c}$ as the initial input of decoder.
First, we adopt a self-attention layer to capture the mutual influence among $n$ candidate ads, producing the set-level augmented embedding $\boldsymbol{h_i}$ for each ad $i$:
\begin{equation}
    \label{equ: encoder self attention}
  [\boldsymbol{h_1}, \boldsymbol{h_2}, \cdots, \boldsymbol{h_n}] = self\operatorname{-}attention([\boldsymbol{e_1}, \boldsymbol{e_2}, \cdots, \boldsymbol{e_n}]),
\end{equation}
where $\boldsymbol{e_i}$ denotes the embedding of ad $a_i$.
Since the encoder processes an unordered set of ads,
the above attention layer excludes positional encoding, and we perform sum pooling on $\boldsymbol{h_i}, i\in [n]$, ensuring that changing the permutation of input ads retains the same output context embedding $\boldsymbol{c}$:
\begin{equation*}
  \boldsymbol{c} = MLP([\sum_{i=1}^n \boldsymbol{h_i}; \boldsymbol{h_u}]),
\end{equation*}
where $\boldsymbol{h_u}$ denotes the user representation.

\textbf{Permutation-equivariant autoregressive decoder.}
To address the limitations of the ``allocation-after-prediction'' paradigm in auction design, we introduce an autoregressive decoder. This model efficiently captures joint distributions over the output ad sequences \cite{autoregressiveCV} to generate the ad sequence. Given the context embedding $\boldsymbol{c}$, the probabilistic generative model learns the conditional probability $p(\boldsymbol{A_m}|\boldsymbol{c})$ for each ad sequence $\boldsymbol{A_m}$. During inference, the model selects a specific allocation from its output space based on this conditional probability.
In the multi-slot auction setting, where $\boldsymbol{A_m}$ consists of $k$ ads: $a_{A_{m_1}}, a_{A_{m_2}}, \cdots, a_{A_{m_k}}$, and $a_{A_{m_i}}$ are not independent, the autoregressive decoder models the joint output distribution using the product rule:
\begin{equation*}
  \begin{aligned}
    p(a_{A_{m_1}}, a_{A_{m_2}}, \cdots, a_{A_{m_k}}|\boldsymbol{c}) = \ &p(a_{A_{m_1}}|\boldsymbol{c})  p(a_{A_{m_2}}|\boldsymbol{c}, a_{A_{m_1}})  \cdots \\ &p(a_{A_{m_k}}|\boldsymbol{c},a_{A_{m_1}}, a_{A_{m_2}}, \cdots, a_{A_{m_{k-1}}}).
  \end{aligned}
\end{equation*}

We further employ the GRU cell \cite{GRU} to model the conditional probability $p(a_i|\boldsymbol{c}, a_{A_{m_1}}, \cdots, a_{A_{m_{t-1}}})$ for each candidate ad $a_i$ at slot $t\in [k] (a_{A_{m_0}}=\varnothing$).
At the beginning of decoding, the context embedding $\boldsymbol{c}$ initializes the hidden state of GRU, and a special token representing "START" is fed into GRU as the initial input. 
Iteratively, the $t$-th GRU cell is formulated as: 
\begin{equation*}
  \boldsymbol{s_{t}}=GRU(\boldsymbol{s_{t-1}}, \boldsymbol{h_{A_{t-1}}}),\ t = 1,2, \cdots,k,
\end{equation*}
where $\boldsymbol{h_{A_{t-1}}}$ denotes the encoded representation of the ad chosen in slot $t-1$ ($\boldsymbol{h_{A_{0}}}=START$), and $\boldsymbol{s_{t-1}}$ denotes the preceding contextual information ($\boldsymbol{s_0}=\boldsymbol{c}$). 
% For brevity, we omit the GRU's internal implementation details.
Consequently, given the state $\boldsymbol{s_{t}}$ at slot $t$, we obtain the allocation probability $z_i^{t}$ of each candidate ad $a_i$:
\begin{equation}
  \label{equ: allocation probability}
  z_i^{t} = softmax([MLP([\boldsymbol{s_{t}}; \boldsymbol{h_i}])+e^w \tilde{\phi}_i]_{i=1}^n)_i,
\end{equation}
where $w$ is a learnable parameter such that $e^{w}$ remains positive, ensuring higher virtual value leads to a greater allocation probability, aligning with the objective of maximizing virtual welfare.
Ads allocated prior to slot $t$ are masked, and a sampling strategy based on $\boldsymbol{z^t}$ determines the ad allocated to slot $t$. This sampling occurs only during training to explore diverse sequence generation strategies.
During inference, the ad with the highest value in $\boldsymbol{z^t}$ is selected.
This selection process is repeated $k$ times to generate an ad sequence of length $k$, denoted as $\boldsymbol{A}=[a_{A_1}, a_{A_2}, \cdots, a_{A_k}]$.

Moreover, since the MLPs and GRU cells in Generator operate on each state-ad pair, and the encoder satisfies permutation invariance,
the decoder exhibits \textit{permutation equivariance}.

\subsection{Evaluator: Permutation-aware Prediction Module}
\label{section: evaluator}
The goal of Evaluator is to predict the permutation-aware CTR $\Theta(\boldsymbol{x_i}; \boldsymbol{A}, \boldsymbol{X}, \boldsymbol{u})$ for each ad $a_{A_i}$  in the ad allocation $\boldsymbol{A}$.
Evaluator takes the allocation embedding $\boldsymbol{H_A}=[\boldsymbol{h}_{A_1}, \boldsymbol{h}_{A_2}, \cdots, \boldsymbol{h}_{A_k}]$, user embedding $\boldsymbol{e_u}$ and point-wise CTR $\boldsymbol{\alpha_A} \in [0,1]^k$ as input, and outputs the permutation-aware CTR.
Each embedding $\boldsymbol{h}_{A_i}$ of ad $a_{A_i}$ is derived from the encoder of Generator, as defined in Equation (\ref{equ: encoder self attention}).
Since CGA operates at the end of the three-stage advertising system (matching, prediction and auction), to fully utilize the user interests captured in the preceding stages, Evaluator leverages the point-wise CTR $\boldsymbol{\alpha_{A}}$ output from the prediction stage, and constructs a Permutation-level Ad Encoder to estimate a personalized externality-aware calibration vector $\boldsymbol{\gamma_A} \in (0,2)^{k}$. These two vectors are then element-wise multiplied to obtain the permutation-aware CTR: $\Theta_{\boldsymbol{A}}=min(\boldsymbol{\alpha_{A}} \odot \boldsymbol{\gamma_A}, \boldsymbol{1})$.
 
Specifically, we adopt a Bi-LSTM \cite{bilstm} layer and a multi-head self-attention \cite{attention} layer to encode the ad sequence embedding $\boldsymbol{H_A}$, where Bi-LSTM effectively captures bidirectional sequence information, while self-attention efficiently captures interactions between distantly positioned ads within the sequence.
Formally, the sequential representation of self-attention layer is defined as: 
\begin{equation*}
  \boldsymbol{H_A^s} = softmax\left(\frac{\boldsymbol{Q_A} \boldsymbol{K_A}^T}{\sqrt{d}}\right) \boldsymbol{V_A},
\end{equation*}
where $d$ denotes the dimension of embeddings, and $\boldsymbol{Q_A}, \boldsymbol{K_A}, \boldsymbol{V_A}$ represents the query, key and value matrices, which is transformed linearly from the sum of allocation embedding $\boldsymbol{H_A}$ and positional encoding $\boldsymbol{PE_{A}}$. Our employed positional encoding mechanism adheres to the sinusoidal version in \cite{attention}, endowing the self-attention layer with the capacity to discern the permutation information.

Next, we obtain the forward output state $\boldsymbol{H_A^f}$ and backforward output state $\boldsymbol{H_A^b}$ of $\boldsymbol{H_A}$ through Bi-LSTM layer. 
\begin{equation*}
  \begin{aligned}
    \boldsymbol{H_A^f} &= Bi\operatorname{-}LSTM_{forward}(\boldsymbol{H_A}) \\
    \boldsymbol{H_A^b} &= Bi\operatorname{-}LSTM_{backward}(\boldsymbol{H_A}).
  \end{aligned}
\end{equation*}
For brevity, we omit Bi-LSTM's implementation details, including the input gates, forget gates, output gates and cell states.

Subsequently, all sequential representations are concatenated with the user representation $\boldsymbol{h_u}$, and fed into a Multilayer Perceptron (MLP) to obtain the externality-aware calibration vector:
\begin{equation*}
  \boldsymbol{\gamma_A} = 2\sigma (r(r([\boldsymbol{H_A^s}; \boldsymbol{H_A^f}; \boldsymbol{H_A^b}; \boldsymbol{h_u}]))),
\end{equation*}
where $\sigma(\cdot)$ and $r(\cdot)$ denote the fully connected layer with sigmoidal and ReLU activation functions, respectively.

Finally, it should be noted that CGA allows any Evaluator model that captures sequential context to improve the G-E framework. In this work, we present an efficient implementation from our practice.

\subsection{Payment Computation}
According to Corollary \ref{corollary: optimal solution}, the optimal payment is defined in Definition \ref{def: optimal mechanism}, where the integral term can be rewritten as an expectation:
\begin{equation*}
  \label{equ: integral payment}
  \int_0^{b_i}\Theta_i(t, \boldsymbol{b_{-i}};\boldsymbol{X}, \boldsymbol{u})dt= b_i E_{t_i\sim U[0,b_i]}[\Theta_i(t_i, \boldsymbol{b_{-i}};\boldsymbol{X}, \boldsymbol{u})],
\end{equation*}
which can be approximated using Monte Carlo sampling.
However, note that $\Theta_i(t_i, \boldsymbol{b_{-i}};\boldsymbol{X}, \boldsymbol{u})=\Theta(\boldsymbol{x_i}; \mathcal{A}(t_i, \boldsymbol{b_{-i}}; \boldsymbol{X}, \boldsymbol{u}), \boldsymbol{X}, \boldsymbol{u})$. 
Processing each sample requires invoking Generator $\mathcal{A}$ and Evaluator $\Theta$, resulting in high computational costs during inference. Reducing the number of samples can mitigate this but increases payment variance, directly affecting platform revenue and advertiser utility.
 
Motivated by the successful application of neural networks to effectively approximate the optimal mechanism in multi-item auction setting in \cite{dutting2019optimal}, we introduce \textit{PaymentNet} to learn the optimal payment rule.
Specifically, the inputs include the allocation embedding $\boldsymbol{H_A} \in \mathbb{R}^{k\times d}$, the self-exclusion bidding profile $\boldsymbol{B^-}=[\boldsymbol{b_{-A_1}}, \boldsymbol{b_{-A_2}}, \cdots, \boldsymbol{b_{-A_k}}] \in \mathbb{R}^{k\times (k-1)}$, and the expected value profile $\boldsymbol{Z} \cdot \boldsymbol{\Theta} \in \mathbb{R}^{k\times 1}$, where $\boldsymbol{Z}=[z_{A_1}^1, z_{A_2}^2, \cdots, z_{A_k}^k]$ denotes the allocation probability defined in Equation (\ref{equ: allocation probability}) and $\boldsymbol{\Theta}=[\Theta_{A_1}, \Theta_{A_2}, \cdots, \Theta_{A_k}]$ denotes the permutation-aware CTR estimated by Evaluator.
Moreover, to satisfy IR constraint, as defined in Definition \ref{IR}, PaymentNet employs a sigmoidal activation function to compute the payment rate $\boldsymbol{\tilde{p}}\in [0,1]^k$, and subsequently outputs the payment $\boldsymbol{p}=\boldsymbol{\tilde{p}} \odot \boldsymbol{b}$.
Formally, the payment rate is defined as:
\begin{equation*}
  \boldsymbol{\tilde{p}} = \sigma(r(r([\boldsymbol{H_A}; \boldsymbol{B^-}; \boldsymbol{Z}\cdot\boldsymbol{\Theta}]))),
\end{equation*}
where $\sigma(\cdot)$ and $r(\cdot)$ denote the fully connected layer with sigmoidal and ReLU activation functions, respectively.

\section{Optimization and Training}
According to Corollary \ref{corollary: optimal solution}, the optimal allocation rule only requires maximizing virtual welfare and is independent of the payment rule. Therefore, we decouple the optimization of G-E framework and PaymentNet, which maintains the optimality of CGA.

\subsection{Optimization of G-E Framework}
In the G-E framework, as Evaluator captures permutation-level externalities and guides Generator to obtain the optimal allocation, we first train Evaluator $\Theta$ to convergence using list-wise ad click data.
Each sample $l \in \mathcal{D}$ is an ad sequence of length $k$ exposed to a user, with the label $y_i^l \in \{0,1\}$ indicating whether the user clicks on the $i$-th ad. The binary cross-entropy loss is defined as:
\begin{equation}
  \label{equ: evaluator loss}
  \mathcal{L}_{E} = - \frac{1}{|\mathcal{D}|} \sum_{l \in \mathcal{D}} \sum_{i=1}^{k} \big( y_i^l \log \theta_i^l + (1-y_i^l) \log (1-\theta_i^l) \big),
\end{equation}
where $\theta_i^l$ denotes the permutation-aware CTR of the $i$-th ad in the ad sequence $l$ as predicted by Evaluator.

After training Evaluator $\Theta$ to convergence, we freeze its parameters and train Generator using policy gradient with rewards from Evaluator.
At each slot $t$, the contextual information $\boldsymbol{s_t}$ serves as the state, and the allocation probability $z_i^{t}$ output by Generator serves as the action.
Since $\boldsymbol{s_t}$ only encodes the preceding context, to capture the bi-directional contextual information, we use the well-trained Evaluator to estimate the permutation-aware reward of each candidate ad within the complete ad sequence $\boldsymbol{A}$.
Similar to \cite{GRN}, we decompose this permutation-aware reward into two parts.

\textbf{Self-Reward.} 
Aligning with the objective of optimal allocation, we use Evaluator $\Theta$ to estimate the expected virtual welfare of each selected ad $a_{A_i}$, termed as self-reward:
\begin{equation}
  \label{equ: self reward}
  r^{self}_{A_i} = \tilde{\phi_i} \cdot \Theta_i(\boldsymbol{b};\boldsymbol{A},\boldsymbol{u}).
\end{equation}

\textbf{External-Reward.}
Each selected ad not only contributes its reward but also affects the CTR of other ads due to the permutation-level externalities. 
Similar to the marginal contribution applied in classical VCG mechanism \cite{VCG}, we model this external effect as external reward:
\begin{equation}
  \label{equ: external reward}
  r^{external}_{A_i} = \sum_{j \in \boldsymbol{A_{-i}}} \tilde{\phi_j} \Theta_j(\boldsymbol{b};\boldsymbol{A},\boldsymbol{u}) - \sum_{j \in \boldsymbol{A_{-i}}} \tilde{\phi_j} \Theta_j(\boldsymbol{b_{-i}};\boldsymbol{A_{-i}},\boldsymbol{u}),
\end{equation}
where $\boldsymbol{A_{-i}}$ denotes the ad sequence excluding $a_{A_i}$.

Combing the above two rewards, we obtain the permutation-aware reward of selecting $a_{A_i}$, defined as:
\begin{equation*}
  \label{equ: permutation-aware reward}
  \begin{aligned}
    r_{A_i} &= r^{self}_{A_i} + r^{external}_{A_i} \\
            &= \tilde{\phi_i} \Theta_i(\boldsymbol{b};\boldsymbol{A},\boldsymbol{u}) + \sum_{j \in \boldsymbol{A_{-i}}} \tilde{\phi_j} \Theta_j(\boldsymbol{b};\boldsymbol{A},\boldsymbol{u}) - \sum_{j \in \boldsymbol{A_{-i}}} \tilde{\phi_j} \Theta_j(\boldsymbol{b_{-i}};\boldsymbol{A_{-i}},\boldsymbol{u}) \\
            &= \sum_{j \in \boldsymbol{A}} \tilde{\phi_j} \Theta_j(\boldsymbol{b};\boldsymbol{A},\boldsymbol{u}) - \sum_{j \in \boldsymbol{A_{-i}}} \tilde{\phi_j} \Theta_j(\boldsymbol{b_{-i}};\boldsymbol{A_{-i}},\boldsymbol{u}).
  \end{aligned}
\end{equation*}
Finally, the loss function of Generator is defined as:
\begin{equation}
  \label{equ: generator loss}
  \mathcal{L}_{G} = - \frac{1}{|\mathcal{D}|} \sum_{s \in \mathcal{D}} \sum_{i \in k} r_{A_i^s} \log z_{A_i^s},
\end{equation}
where $s$ denotes a sample from dataset $\mathcal{D}$, representing the candidate ad set for a request, $A_i^s$ denotes the $i$-th ad of the ad allocation output by Generator based on the input ad set $s$, and $z_{A_i^s}$ denotes the allocation probability in Equation (\ref{equ: allocation probability}).

\subsection{Optimization of PaymentNet}
After training Generator $\mathcal{A}$ and Evaluator $\Theta$ to convergence, we freeze their parameters and then apply the augmented Lagrangian method to solve the constrained optimization problem (\ref{equ: learning-based problem formulation}) to optimize PaymentNet.
The Lagrangian function augmented with a quadratic penalty term for violating the IC constraint is defined as:
\begin{equation}
  \label{equ: Lagrangian}
  \mathcal{L}_P = - \frac{1}{|\mathcal{D}|} \sum_{s \in \mathcal{D}} \big( \sum_{i \in k}\mathcal{P}_i(\boldsymbol{A}^s) \Theta_i(\boldsymbol{A}^s) - \sum_{i \in k} \lambda_i\widehat{rgt}_i^s - \frac{\rho}{2}\sum_{i \in k}(\widehat{rgt}_i^s)^2  \big),
\end{equation}
where $\boldsymbol{A}^s$ denotes the allocation output by the freezed Generator $\mathcal{A}$ with the input of ad set $s$, $\lambda_i$ denotes Lagrange multipliers, and $\rho >0$ denotes the hyperparameter for the IC penalty term.

Similar to \cite{dutting2019optimal}, the optimization process based on Lagrangian function (\ref{equ: Lagrangian}) includes the iteration of
(\textbf{\romannumeral1}) updating PaymentNet: $\boldsymbol{w_\mathcal{P}^{new}} = argmin_{w_\mathcal{P}} \mathcal{L}_P(\boldsymbol{w_\mathcal{P}^{old}}; \boldsymbol{\lambda^{old}})$,
and (\textbf{\romannumeral2}) updating the Lagrange multipliers: $\boldsymbol{\lambda^{new}} = \boldsymbol{\lambda^{old}} + \rho\  \boldsymbol{\widehat{rgt}}(\boldsymbol{w_\mathcal{P}^{new}})$.
Note that problem (\ref{equ: learning-based problem formulation}) is non-convex. Thus, the above Lagrangian method is not guaranteed to converge to the global optimum. However, the experimental results indicate that the optimized CGA can approximate the optimal revenue with minimal regret, as discussed in Section \ref{sec: performance comparison}.

%% file: sections/Experiments.tex
\section{Experiments}

In this section, we conduct offline experiments and online A/B tests to evaluate the effectiveness of CGA, and mainly focus on the following questions:

$\bullet$ \textbf{RQ1}: How does CGA perform in terms of key metrics like platform revenue and CTR, compared with existing ad auctions?

$\bullet$ \textbf{RQ2}: What are the effects of modeling permutation-level externalities compared to set-level and W/O externalities?

$\bullet$ \textbf{RQ3}: How do various designs of CGA affect its performance?

$\bullet$ \textbf{RQ4}: How does our proposed CGA perform in the real-world ad auction scenarios with efficient deployment?

\subsection{Experiment Setup}

\subsubsection{Dataset}
In offline experiments, we evaluate the performance of CGA using real logs collected from a leading e-commerce platform \textit{Taobao} during January 2024. The training set comprises 500,000 randomly selected auctions from January 20 to January 23, involving 1,100,875 unique advertisers while the test set comprises 100,000 randomly selected auctions from January 25, involving 452,671 unique advertisers.
For each auction sample in the dataset, approximately 30 advertisers are selected by ad system as candidates for the auction stage, where they submit their bids to compete for slots. We set the number of slots $k=3$ in offline experiments.

\subsubsection{Baseline Methods}
\label{sec: baseline}

We compare CGA with representative auction mechanisms.
The baseline methods are categorized into three groups based on the granularity of externality modeling.

\noindent \textbf{W/O externalities:} 

$\bullet$ \textbf{GSP}: 
GSP ranks ads based on the product of bids and predicted CTR, without modeling externalities.

\noindent \textbf{Set-level externalities}:

$\bullet$ \textbf{DNA} \cite{dna}: 
Building on GSP, DNA predicts each ad's rankscore by modeling set-level externalities of candidate ads and ranks accordingly for allocation, without considering the mutual influence among ads in the ad list.

$\bullet$ \textbf{SW-VCG} \cite{li2023learning}:
SW-VCG formalizes multi-slot auction design as a maximum weighted bipartite matching problem between ads and slots, which captures only set-level externalities of candidate ads to estimate edge weights.

\noindent \textbf{Permutation-level externalities}: 

$\bullet$ VCG Auction \cite{VCG} with Permutation-level Externalities (\textbf{VCG}): 
VCG selects the ad permutation that maximizes social welfare for allocation and combines this with a payment rule to satisfy IC. We evaluate all permutations of the candidate ads with CGA's Evaluator, enabling VCG to capture permutation-level externalities.

$\bullet$ \textbf{EdgeNet} \cite{edgenet}: EdgeNet utilizes Transformer to replace DNA's set encoder and employs a greedy strategy to sequentially allocate ads, modeling partial permutation-level externalities.

$\bullet$ \textbf{Optimal}: 
To evaluate CGA's approximation to the theoretical upper bound, according to Corollary \ref{corollary: optimal solution}, we construct this baseline by traversing all permutations to maximize virtual welfare and using Monte Carlo sampling to calculate the payment.

\begin{table*}[htbp]
    \centering
    \caption{Performance comparison for key metrics. Lift percentage indicates the relative change of baseline methods over CGA.}
    \label{table: performance}
    \resizebox{0.8\linewidth}{!}{
        \begin{tabular}{c|c|c|ccc}
            \hline
            Value Distribution                 & Externalities                      & Model        & RPM               & CTR               & $\Psi$         \\ \hline \hline
            \multirow{7}{*}{Uniform}     & W/O Externalities                  & GSP          & 1141.40 (-9.7\%)  & 0.04359 (-7.1\% ) & 14.7\%         \\ \cline{2-3}
                                         & \multirow{2}{*}{Set-level}         & DNA          & 1195.75 (-5.4\%)  & 0.04537 (-3.3\%)  & 8.4\%          \\
                                         &                                    & SW-VCG       & 1190.70 (-5.8\%)  & 0.04528 (-3.5\%)  & 5.6\%          \\ \cline{2-3}
                                         & \multirow{4}{*}{Permutation-level} & EdgeNet      & 1205.87 (-4.6\%)  & 0.04561 (-2.8\%)  & 4.9\%          \\
                                         &                                    & VCG          & 1085.78 (-14.1\%) & 0.04791 (+2.1\%)  & 0.0            \\
                                         &                                    & Optimal      & 1318.36 (+4.3\%)  & 0.04772 (+1.7\%)  & 0.0            \\
                                         &                                    & \textbf{CGA} & \textbf{1264.01}  & \textbf{0.04692}  & \textbf{2.1\%} \\ \hline
            \multirow{7}{*}{Exponential} & W/O Externalities                  & GSP          & 1179.22 (-12.2\%) & 0.04050 (-9.8\%)  & 18.3\%         \\ \cline{2-3}
                                         & \multirow{2}{*}{Set-level}         & DNA          & 1257.11 (-6.4\%)  & 0.04283 (-4.6\%)  & 10.1\%         \\
                                         &                                    & SW-VCG       & 1247.71 (-7.1\%)  & 0.04266 (-5.0\%)  & 7.4\%          \\ \cline{2-3}
                                         & \multirow{4}{*}{Permutation-level} & EdgeNet      & 1266.52 (-5.7\%)  & 0.04315 (-3.9\%)  & 6.7\%          \\
                                         &                                    & VCG          & 1321.85 (-15.8\%) & 0.04602 (+2.5\%)  & 0.0            \\
                                         &                                    & Optimal      & 1415.60 (+5.4\%)  & 0.04589 (+2.2\%)  & 0.0            \\
                                         &                                    & \textbf{CGA} & \textbf{1343.07}  & \textbf{0.04490}  & \textbf{3.7\%} \\ \hline
            \end{tabular}
    }
\end{table*}

\subsubsection{Performance Metrics}
We consider the following metrics to measure platform revenue, user experience, and ex-post regret of advertisers, respectively.
 
$\bullet$ Revenue Per Mille: $\text{RPM} = \frac{\sum click \times payment}{\sum impression} \times 1000.$

$\bullet$ Click-Through Rate: $\text{CTR} = \frac{\sum click}{\sum impression}.$

$\bullet$ IC metric: $\Psi = \frac{1}{|\mathcal{D}|} \sum_{s \in \mathcal{D}} \sum_{i \in k} \frac{\widehat{rgt}_i^s}{u_i(v_i^s, \boldsymbol{b^s}; \boldsymbol{X^s}, \boldsymbol{u^s})},$
where $\widehat{rgt}_i^s$ is defined in Equation (\ref{equ: empirical ex-post regret}).
This is a common metric for IC testing of ad auction mechanisms \cite{ICmetric, dna, edgenet,nma}, which measures the relative increase in utility that advertisers can obtain by manipulating their bids. 
Consistent with the evaluation process in \cite{twostageauction}, we conduct counterfactual perturbation on each advertiser’s bid by replacing $b_i$ with $\alpha \times b_i$, where $\alpha \in \{0.2 \times j \mid j=1,2, \cdots, 10\}$.

\subsubsection{Implementation} In CGA, we set the embedding size of features as 8. To capture richer information from different representation subspaces, a multi-head attention mechanism with 4 attention heads is used in all attention layers. The externality-aware calibration vector $\boldsymbol{\gamma_A}$ and the payment rate $\boldsymbol{\tilde{p}}$ are derived using a MLP with hidden layers of sizes 128 and 32. CGA is trained with the Adam optimizer at a learning rate of 1e-3 and a batch size of 512. All learning-based baseline methods have hyperparameters tuned through grid search, exploring learning rates of \{1e-5, 1e-4, 1e-3, 1e-2\}, batch sizes of \{256, 512, 1024, 2048\}, and hidden sizes of \{8,16,32,64\}.

\subsection{Offline Comparison (RQ1 \& RQ2)}
\label{sec: performance comparison}
CGA and baseline-Optimal involve the ironed virtual value function,
which requires knowledge of each advertiser's value distribution $f_i(\cdot)$.
Estimating $f_i(\cdot)$ from historical data is well-covered in existing research \cite{jiang2005estimating, noti2021bid, ostrovsky2023reserve, cherapanamjeri2022estimation}.
This paper does not delve into this estimation as CGA focuses on employing generative models to capture permutation-level externalities, independent of distribution estimation techniques.
To avoid biases from distribution estimation methods in comparative experiments, following \cite{li2023learning}, we preset the conditional distribution of each advertiser's value based on their feature vectors as \textit{uniform} and \textit{exponential distributions}, respectively, and regenerate the bids accordingly.
For SW-VCG, since its predicted ad score $Scr_i$ aims to fit the virtual value, we directly set $Scr_i$ to $\tilde{\phi}_i$ for comparison.

The experimental results under the above two distributions are presented in Table \ref{table: performance}. Our key observations are:

(1) As externality modeling advances from W/O externalities to set-level and finally to permutation-level, auction performance improves across three metrics (higher RPM and CTR with lower regret)\footnote{VCG optimizes social welfare at the expense of revenue, independent of externalities.}. 
This highlights the importance of modeling fine-grained externalities in auction design (\textbf{RQ2}).
To further investigate the impact of externalities, we conduct comparative experiments on the CTR prediction task. The results are shown in Table \ref{table: externalities}, where "W/O Externalities" indicates the use of point-wise pCTR and "Set-level" refers to using the Set-level Ad Encoder, as described in Section \ref{section: evaluator}, to replace Evaluator's permutation-level ad encoder. The results demonstrate that Evaluator with permutation-level externalities improves the accuracy of predictive values, explaining the improvement in auction metrics.

(2) Compared to baseline-Optimal based on enumeration, CGA achieves approximately 95\% revenue maximization and negligible ex-post regret. This suggests that CGA enables efficient ad allocation and closely approximates the optimal auction mechanism.

\begin{table}[htbp]
    \centering
    \caption{Performance comparison of externalities modeling.}
    \label{table: externalities}
    \resizebox{0.9\linewidth}{!}{
        \begin{tabular}{c|cc}
            \hline
                ~ & Logloss & AUC \\ \hline \hline
                W/O Externalities & 0.2311 & 0.7366 \\ 
                Set-level Externalities  & 0.2286 & 0.7391 \\ 
                Permutation-level Evaluator $\Theta$ & \textbf{0.2237} & \textbf{0.7476} \\ \hline
        \end{tabular}
    }
\end{table}

\subsection{Ablation Study (RQ3)}
\label{section: ablation}
To verify the effectiveness of CGA's various design considerations, we construct the following variants:

$\bullet$ CGA-$\Theta$ removes the Evaluator $\Theta$ and uses point-wise CTR instead to evaluate each ad allocation's virtual welfare.

$\bullet$ CGA (end2end) directly trains Generator and PaymentNet using $\mathcal{L}_P$ from Equation (\ref{equ: Lagrangian}). In contrast, CGA first trains Generator to convergence, freezes its parameters, and then trains PaymentNet.

$\bullet$ CGA-$r^{self}$ removes the self reward $r^{self}$, defined in Equation (\ref{equ: self reward}), in the loss function $\mathcal{L}_G$ of Generator. 

$\bullet$ CGA-$r^{external}$ removes the external reward $r^{external}$, defined in Equation (\ref{equ: external reward}), in the loss function $\mathcal{L}_G$ of Generator. 

$\bullet$ CGA-$\tilde{\phi}$ replaces virtual value $\tilde{\phi_i}$ with $b_i$ in $\mathcal{L}_G$.

The results are shown in Table \ref{table: ablation}, from which we observe:

(1) CGA-$\Theta$ performs worse than CGA, indicating that the Generator alone cannot fully capture permutation-level externalities, as the autoregressive model perceives only preceding context. The G-E framework helps distill complete sequential knowledge into Generator via policy gradient.

(2) CGA (end2end) performs worse. One likely reason is that CGA uses PaymentNet to learn the optimal payment, so the revenue gradient must pass through PaymentNet before reaching Generator while conducting end-to-end training, complicating convergence to the optimal allocation. Furthermore, Corollary \ref{corollary: optimal solution} ensures that decoupled optimization maintains optimality, thus allowing Generator to focus solely on maximizing virtual welfare.

(3) CGA-$r^{self}$ performs the worst due to the absence of each selected ad's own value increment. CGA outperforms CGA-$r^{external}$ because it uses external rewards to model the impact of each selected ad on the final allocation list.

(4) Estimating value distribution $f_i(\cdot)$ requires empirical assumptions about bidding strategies, since Ad platforms can only observe bids, causing bias in the estimated $\hat{f_i}(\cdot)$.
To assess the impact of removing value distribution on CGA's performance, we construct the variant CGA-$\tilde{\phi}$. As shown in Table \ref{table: ablation}, CGA-$\tilde{\phi}$ shows only a minor decline in revenue (2.6\%) and CTR (1.9\%) compared to CGA. This small drop may be due to CGA's modeling of externalities, which captures inter-ad correlations and partially reflects the missing value distribution information through other advertisers' bids.

\begin{table}[htbp]
    \centering
    \caption{Ablation study of CGA.}
    \label{table: ablation}
    \resizebox{\linewidth}{!}{
    \begin{tabular}{l|ccc}
    \hline
        Model & RPM & CTR & $\Psi$ \\ \hline \hline
        \textbf{CGA} & \textbf{1264.01} & \textbf{0.04692} & \textbf{2.1\%}  \\ 
        CGA-$\Theta$ & 1214.71 (-3.9\%) & 0.04579 (-2.4\%) & 3.7\%  \\ 
        CGA (end2end) & 1181.85 (-6.5\%)  & 0.04504 (-4.0\%) & 10.5\%  \\ 
        CGA-$r^{self}$ & 1122.44 (-11.2\%)   & 0.04302 (-8.3\%) & 18.9\% \\ 
        CGA-$r^{external}$ &  1210.92 (-4.2\%) & 0.04575 (-2.5\%) & 3.4\%  \\ 
        CGA-$\tilde{\phi}$ & 1231.15 (-2.6\%) & 0.04603 (-1.9\%) & 2.8\% \\ \hline
    \end{tabular}
    }
\end{table}

\subsection{Online A/B Test (RQ4)}
\label{section: online}
 
To verify CGA's effectiveness in the real-world, we compare CGA with the fully deployed DNA in Taobao advertising system through online A/B tests.
Table \ref{table: online results} presents the results of online A/B testing conducted from August 19 to August 25, 2023, utilizing 2\% of total production traffic.
CGA achieves a 3.2\% improvement in RPM with only a 3 ms average increase (1.6\% relatively) in online response time (RT) per request, suggesting that CGA can efficiently explore the allocation space via generative models and boost platform revenue.
Moreover, the heightened return on investment (ROI) for advertisers indicates that CGA's revenue enhancement results not from inflated payments but from optimized ad allocation by capturing permutation-level externalities, as evidenced by significant CTR and gross merchandise volume (GMV) improvements.

\begin{table}[htbp]
    \centering
    \caption{Experimental results from Online A/B tests.}
    \label{table: online results}
    \resizebox{\linewidth}{!}{
    \begin{tabular}{l|ccccc}
    \hline
        Relative change in metrics & RPM & CTR & GMV & ROI & RT \\ \hline \hline
        \textbf{CGA} over baseline-DNA & +3.2\% & +1.4\% & +6.4\% & +3.5\% & +1.6\% \\ \hline
    \end{tabular}
    }
\end{table}

%% file: sections/Related_work.tex
\section{Related Work}
GSP \cite{edelman2007internet} and its variants, like uGSP \cite{ugsp}, are widely used in online advertising due to their interpretability and high revenue guarantee, but they do not consider the impact of other ads on user clicks, neglecting externalities \cite{externalities_auction_1, externalities_auction_2}, leading to suboptimal performance.

Advances in computing have led researchers to explore learning-based auctions \cite{zhang2021survey}.
DeepGSP \cite{DeepGSP} and DNA \cite{dna} extend GSP using online feedback for end-to-end learning. However, DNA's rankscore prediction faces the evaluation-before-ranking problem \cite{evaluaton-before-rank}, restricting its scope to set-level externalities and yielding suboptimal allocations.
SW-VCG \cite{li2023learning} separates optimal auction design into designing monotone score functions and solving ad-slot maximum bipartite matching. However, SW-VCG still overlooks the mutual influence of ads in the final sequence, as the edge weights in the bipartite graph cannot be predetermined considering permutation-level externalities.
Following VCG, NMA \cite{nma} proposed an enumeration-based framework to select the optimal allocation. While globally optimal, its high computational complexity makes it impractical for real-time online inference \cite{PIER}. EdgeNet \cite{edgenet} employs a PointerNet-based structure for greedy ad allocation but ignores the impact of succeeding ads, remaining insufficient for achieving optimal results.

Studies of reranking in RS parallels externality modeling by leveraging contextual information to optimize item sequences. Reranking methods can be divided into one-stage and two-stage approaches \cite{ren2024non}.
One-stage methods \cite{OneStageRerank_1, pei2019personalized} estimate refined scores for each item within the initial list and rerank them greedily. These methods encounter the evaluation-before-ranking problem similar to DNA: reranking alters the permutation, introducing different mutual influences.
Two-stage methods \cite{GRN,PIER, ren2024non} typically employ a G-E framework. The Generator produces multiple feasible sequences, while the Evaluator selects the optimal sequence based on the estimated list value. This approach enables comprehensive exploration of the permutation space \cite{ren2024non}, which provides valuable insights for the implementation of CGA.
While the G-E framework effectively captures permutation-level externalities, it lacks the capacity to express key economic constraints such as IC and fails to directly optimize platform revenue.
Our theoretical results decouple the allocation and payment in optimal auctions with permutation-level externalities. This allows the allocation to employ a general G-E framework focused solely on maximizing expected virtual welfare, while the optimal payment rule is learned through differentiable ex-post regret. 
Consequently, CGA can be viewed as a unified framework bridging reranking in RS and ad auction theory.

%% file: sections/Conclusion.tex
\section{CONCLUSION}
This paper proposes the Contextual Generative Auction (CGA), designed to incorporate permutation-level externalities in online multi-slot ad auctions. 
Our primary theoretical results demonstrate that the classic Myerson Auction maintains its optimality when adapted to permutation-level externalities.
This insight drives the design of the CGA framework, which decouples the optimization of allocation and payment.
Specifically, we develop an autoregressive generative model with G-E learning paradigm to optimize allocation, and learn the optimal payment by quantifying IC constraint into expected ex-post regret.
Extensive offline and online experiments verify the effectiveness of CGA.
Notably, the autoregressive generation process of CGA is not limited to specific generative model and can accommodate various advanced solutions \cite{FLOW, wang2023diffusion}. Future research will extend this contextual generative mechanism to integrate heterogeneous items from different channels.

%% file: sections/Appendix.tex
\appendix
\section{APPENDIX}

\subsection{Proof of Lemma \ref{lemma: monotone allocation}}
\label{appendix: proof of monotone allocation}

% \textit{Proof of Lemma \ref{lemma: monotone allocation}}. 
Denote $\boldsymbol{A^*}$ as the allocation that maximizes the expected virtual welfare (select randomly if there exist multiple optimal allocation), \ie, 
\begin{equation*}
  \boldsymbol{A^*} = \mathcal{A}(b_t, \boldsymbol{b_{-t}}; \boldsymbol{X}, \boldsymbol{u}) =  argmax_{A}\sum_{i=1}^{n}\tilde{\phi}(b_i, F_i) \cdot \Theta(\boldsymbol{x_i}; \boldsymbol{A}, \boldsymbol{X}, \boldsymbol{u}),
\end{equation*}
which can be obtained through enumeration. 
\textit{W.l.o.g.}, suppose $ad_t \in \boldsymbol{A^*}$, and the other ads in $\boldsymbol{A^*}$ constitute the set $\boldsymbol{A_{-t}^*}$, then the virtual welfare of $\boldsymbol{A^*}$ can be written as:
\begin{equation*}
  \tilde{\Phi}(\boldsymbol{A^*}; b_t) = \tilde{\phi}(b_t, F_t)\Theta(\boldsymbol{x_t}; \boldsymbol{\boldsymbol{A^*}}, \boldsymbol{X}, \boldsymbol{u}) + \sum_{j\in \boldsymbol{A_{-t}^*}} \tilde{\phi}(b_j, F_j)\Theta(\boldsymbol{x_j}; \boldsymbol{\boldsymbol{A^*}}, \boldsymbol{X}, \boldsymbol{u}).
\end{equation*}
When the bid of $ad_t$ rises to $b_t'>b_t$, suppose the resulting allocation maximizing virtual welfare is $\boldsymbol{\boldsymbol{A'}}$, \ie, 
\begin{equation*}
  \boldsymbol{\boldsymbol{A'}} = \mathcal{A}(b_t', \boldsymbol{b_{-t}}; \boldsymbol{X}, \boldsymbol{u}).
\end{equation*}

We can directly obtain that $ad_t \in \boldsymbol{\boldsymbol{A'}}$; otherwise, suppose $ad_t \notin \boldsymbol{\boldsymbol{A'}}$, then we have $\tilde{\Phi}(\boldsymbol{\boldsymbol{A'}}; b_t') = \tilde{\Phi}(\boldsymbol{\boldsymbol{A'}}; b_t)$.
Since $\boldsymbol{A^*}$ maximizes virtual welfare when $ad_t$ bids $b_t$ and $\phi(b, F)$ is monotone non-decreasing in $b$ \cite{myerson1981optimal}, it follows that $\tilde{\Phi}(\boldsymbol{\boldsymbol{A'}}; b_t) \leq \tilde{\Phi}(\boldsymbol{A^*}, b_t) \leq \tilde{\Phi}(\boldsymbol{A^*}; b_t')$.
If the first '$\leq$' holds as equality, meaning when $ad_t$ bids $b_t'$, both allocations $\boldsymbol{A'}$ and $\boldsymbol{A^*}$ have equal and maximized virtual welfare, we choose $\boldsymbol{A^*}$ as the allocation result \textit{w.l.o.g}. Thus, when $ad_t$ increases its bid, the allocation remains unchanged, and so does the CTR of $ad_t$, validating the proposition.
Otherwise, $\tilde{\Phi}(\boldsymbol{A'}; b_t) < \tilde{\Phi}(\boldsymbol{A^*}, b_t)$, hence, $\tilde{\Phi}(\boldsymbol{A'}; b_t') < \tilde{\Phi}(\boldsymbol{A^*}; b_t')$, contradicting the definition of $\boldsymbol{A'}$. Therefore, we have $ad_t \in \boldsymbol{A'}$.

Similarly, the virtual welfare of $\boldsymbol{A'}$ can be written as:
\begin{equation*}
  \tilde{\Phi}(\boldsymbol{A'}; b_t') = \tilde{\phi}(b_t', F_t)\Theta(\boldsymbol{x_t}; \boldsymbol{\boldsymbol{A'}}, \boldsymbol{X}, \boldsymbol{u}) + \sum_{j\in \boldsymbol{A_{-t}'}} \tilde{\phi}(b_j, F_j)\Theta(\boldsymbol{x_j}; \boldsymbol{\boldsymbol{A'}}, \boldsymbol{X}, \boldsymbol{u}).
\end{equation*}
Based on the maximized virtual welfare property of $\boldsymbol{A^*}$ and $\boldsymbol{A'}$, we can deduce the following two inequalities:
\begin{equation}
  \label{equ: proof 1}
  \begin{aligned}
    & \tilde{\phi}(b_t, F_t)\Theta(\boldsymbol{x_t}; \boldsymbol{\boldsymbol{A^*}}, \boldsymbol{X}, \boldsymbol{u}) + \sum_{j\in \boldsymbol{A_{-t}^*}} \tilde{\phi}(b_j, F_j)\Theta(\boldsymbol{x_j}; \boldsymbol{\boldsymbol{A^*}}, \boldsymbol{X}, \boldsymbol{u}) \geq \\
  & \tilde{\phi}(b_t, F_t)\Theta(\boldsymbol{x_t}; \boldsymbol{\boldsymbol{A'}}, \boldsymbol{X}, \boldsymbol{u}) + \sum_{j\in \boldsymbol{A_{-t}'}} \tilde{\phi}(b_j, F_j)\Theta(\boldsymbol{x_j}; \boldsymbol{\boldsymbol{A'}}, \boldsymbol{X}, \boldsymbol{u}),
  \end{aligned}
\end{equation}

\begin{equation}
  \label{equ: proof 2}
  \begin{aligned}
    & \tilde{\phi}(b_t', F_t)\Theta(\boldsymbol{x_t}; \boldsymbol{\boldsymbol{A^*}}, \boldsymbol{X}, \boldsymbol{u}) + \sum_{j\in \boldsymbol{A_{-t}^*}} \tilde{\phi}(b_j, F_j)\Theta(\boldsymbol{x_j}; \boldsymbol{\boldsymbol{A^*}}, \boldsymbol{X}, \boldsymbol{u}) \leq \\
  & \tilde{\phi}(b_t', F_t)\Theta(\boldsymbol{x_t}; \boldsymbol{\boldsymbol{A'}}, \boldsymbol{X}, \boldsymbol{u}) + \sum_{j\in \boldsymbol{A_{-t}'}} \tilde{\phi}(b_j, F_j)\Theta(\boldsymbol{x_j}; \boldsymbol{\boldsymbol{A'}}, \boldsymbol{X}, \boldsymbol{u}).
  \end{aligned}
\end{equation}

By subtracting Inequality (\ref{equ: proof 2}) from Inequality (\ref{equ: proof 1}), we obtain:
\begin{equation}
  \begin{aligned}
  & (\tilde{\phi}(b_t, F_t) - \tilde{\phi}(b_t', F_t))\Theta(\boldsymbol{x_t}; \boldsymbol{\boldsymbol{A^*}}, \boldsymbol{X}, \boldsymbol{u}) \geq \\
  & (\tilde{\phi}(b_t, F_t) - \tilde{\phi}(b_t', F_t))\Theta(\boldsymbol{x_t}; \boldsymbol{\boldsymbol{A'}}, \boldsymbol{X}, \boldsymbol{u}).
  \end{aligned}
\end{equation}

Since $\tilde{\phi}(t, F_t)$ is monotone non-decreasing in $t$ and $b_t < b_t'$, we have:
\begin{equation}
  \tilde{\phi}(b_t, F_t) \leq \tilde{\phi}(b_t', F_t).
\end{equation}

$\bullet$ Case 1: $\tilde{\phi}(b_t, F_t) - \tilde{\phi}(b_t', F_t) < 0$. Then we have:
\begin{equation*}
  \Theta(\boldsymbol{x_t}; \boldsymbol{\boldsymbol{A^*}}, \boldsymbol{X}, \boldsymbol{u}) \leq \Theta(\boldsymbol{x_t}; \boldsymbol{\boldsymbol{A'}}, \boldsymbol{X}, \boldsymbol{u}), i.e.,
\end{equation*}
\begin{equation*}
  \Theta(\boldsymbol{x_t}; \mathcal{A}(b_t, \boldsymbol{b_{-t}}; \boldsymbol{X}, \boldsymbol{u}), \boldsymbol{X}, \boldsymbol{u}) \leq \Theta(\boldsymbol{x_t}; \mathcal{A}(b_t', \boldsymbol{b_{-t}}; \boldsymbol{X}), \boldsymbol{X}, \boldsymbol{u}),
\end{equation*}
demonstrating that the allocation rule $\mathcal{A}$ is monotone.

$\bullet$ Case 2: $\tilde{\phi}(b_t, F_t) - \tilde{\phi}(b_t', F_t) = 0$.
By swapping the LHS and RHS of Inequality (\ref{equ: proof 2}) and then adding it to Inequality (\ref{equ: proof 1}), we obtain:
\begin{equation}
  \label{equ: proof 3}
  \begin{aligned}
    & \tilde{\phi}(b_t, F_t)\Theta(\boldsymbol{x_t}; \boldsymbol{\boldsymbol{A^*}}, \boldsymbol{X}, \boldsymbol{u}) + \tilde{\phi}(b_t', F_t)\Theta(\boldsymbol{x_t}; \boldsymbol{\boldsymbol{A'}}, \boldsymbol{X}, \boldsymbol{u}) \geq \\ 
    & \tilde{\phi}(b_t, F_t)\Theta(\boldsymbol{x_t}; \boldsymbol{\boldsymbol{A'}}, \boldsymbol{X}, \boldsymbol{u}) +  \tilde{\phi}(b_t', F_t)\Theta(\boldsymbol{x_t}; \boldsymbol{\boldsymbol{A^*}}, \boldsymbol{X}, \boldsymbol{u}).
  \end{aligned}
\end{equation}
Substituting $\tilde{\phi}(b_t, F_t) = \tilde{\phi}(b_t', F_t)$ into Equation (\ref{equ: proof 3}), we have, the '$\geq$' in Inequality (\ref{equ: proof 3}) holds as equality. 
Therefore, both Inequality (\ref{equ: proof 1}) and Inequality (\ref{equ: proof 2}) hold as equalities, which means that both $\boldsymbol{A^*}$ and $\boldsymbol{A'}$ are the allocations that maximize virtual welfare after $ad_t$ increases her bid to $b_t'$.
Therefore, \textit{w.l.o.g}, we choose $\boldsymbol{A^*}$ as the final allocation result, indicating that the allocation outcome remains unchanged after the increase in $ad_t$'s bid. Consequently, the CTR of $ad_t$ also remains unchanged. Thus, the allocation rule $\mathcal{A}$ is monotone.